\documentclass[fleqn,twoside]{article}
\usepackage{espcrc2}
\usepackage{epsfig}
\usepackage[figuresright]{rotating}

\newcommand{\AmS}{{\protect\the\textfont2
  A\kern-.1667em\lower.5ex\hbox{M}\kern-.125emS}}

\title{Crossover from superfluidity to superconductivity 
       in a system with doping dependent attraction}

\author{V.M. Loktev\address{Bogolyubov Institute for Theoretical 
        Physics, Metrologichna str. 14-b, Kiev, 03143 Ukraine}%
        and
         V. Turkowski\address{CFIF, Instituto Superior Tecnico,
         Av.Rovisco Pais, 1049-001 Lisbon, Portugal}
\thanks{Corresponding author.Tel.: +351-21-8419094; fax: +351-21-8419143.
         {\it E-mail address:} vturk@cfif.ist.utl.pt 
        }
       }
       
\begin{document}

\begin{abstract}
Zero temperature crossover from superfluidity to superconductivity
with carrier density increasing is studied for a two-dimensional
system in the $s$-wave and $d$-wave pairing channels.
It was assumed that the particle attraction correlation length
depends on carrier density $n_{f}$ as $r_0\sim 1/\sqrt{n_f}$. 
Such a dependence was found experimentally
for the radius of magnetic correlations in $La_{2-x}Sr_{x}CuO_{4}$. 
The short range Coulomb repulsion was also taken into account.
It is shown that the behavior of the system with doping is fundamentally
different from the case with $r_0(n_{f})=const$. In particular, similarly
to the $d$-wave case, the crossover in the $s$-channel takes place only 
if the coupling is larger of some minimal value, otherwise the 
Cooper pairing scenario takes place at any small carrier density. 
The relevance of the model to the high-temperature superconductors
 is discussed.

\vspace{1pc}
{\it PACS} : 74.20.-z, 74.62.Dh, 74.72.-h

{\it Keywords} : Symmetry, $s$-Wave, $d$-Wave, Effects of Doping

\end{abstract}

\maketitle

\section{INTRODUCTION}

The problem of the crossover from superfluidity to superconductivity
with charge carrier density or coupling constant changing has a long history 
\cite{Eagles}.
The interest to this phenomena has arisen again after the
discovery of high-temperature superconductors (HTSCs) in 1986
\cite{Micnas,Randeria,Haussmann,Gorbar1,Gorbar2,TMF,denHertog,Andrenacci,Duncan}. 
It was already known upon that time that the superfluidity 
of composite bosons transforms into the superconductivity 
of overlapped Cooper pairs with chemical doping in the case 
of the s-wave pairing. Now, in the
s-wave pairing case the problem is quite well explored
for the 3D systems \cite{Micnas,Randeria,Haussmann} and, particularly
for the quasi-2D case \cite{Gorbar1}. For the 2D case this problem
was studied at $T=0$ ( when a long-range    
superconducting is still possible in a 2D system 
\cite{Mermin})
for the case of local attraction (see, for example 
\cite{Micnas,Randeria,Gorbar2}) and for the phonon-exchange model
\cite{TMF}. Most of these problems are reviewed in \cite{PhysRep}.

Due to a layered structure and the anisotropic symmetry 
of the order parameter in HTSCs, the crossover in the 2D system 
in the $d$-wave pairing channel is of a special interest. 
However, because if its complexity, this case
is not so well understood at present. The d-wave pairing for the case of 
the extended Hubbard model with the nearest neighbor attraction case
was analyzed in \cite{denHertog,Andrenacci}. The crossover
from superfluidity to superconductivity in the $s$-wave and $d$-wave 
pairing channels for a 2D continuum model was studied in paper \cite{Duncan},
where also thermodynamic properties of the system in 
the crossover particle density region were considered.
The authors proposed an interaction potential, which is  attractive 
at distances between particles shorter of some value $r_0$ and longer of $r_1$,
and repulsive due to electron correlations at short distances, $r<r_1$.
It was found, in particular, that there exists a minimal
value of the attractive coupling constant in the $d$-wave pairing channel, 
which gives the crossover from superfluidity to superconductivity
at small carrier densities, i.e. the fermion chemical potential
changes its sign and becomes positive with carrier density growth.

The correlation length $r_0$ was assumed in \cite{Duncan} to be a parameter,
which does not depend on the carrier density $n_f$ per cell. However, 
concerning HTSCs it would be interesting to consider a more realistic
case, when at small carrier densities $r_0\sim 1/\sqrt{n_f}$ 
and the proportionality coefficient is of order of a lattice constant.
 Such a dependence was experimentally observed in HTSC
cuprates for the length of spin-spin correlations, which are believed
to be responsible for the hole attraction in these materials. In particular,
it was found for $La_{2-x}Sr_{x}CuO_{4}$ in underdoped regime,
that the magnetic correlation length decreases with carrier density per cell
according to the dependence $3.8$\AA$/\sqrt{n_f}$ \cite{Thurston}.

In what follows we analyze the possibility of low carrier density crossover 
for a model analogous to \cite {Duncan} with particle repulsion at 
distances $r< r_{1}$ and  attraction at $r_1 < r < r_0$, where, however,
\begin{equation}
 r_0=\frac{a}{\sqrt{n_f}}
\label{R0}
\end{equation}
($a$ is parameter of order of the lattice constant,
and its possible value is discussed in the next Section). 
Obviously, at large carrier densities such that $r_{1}>r_{0}$
the superconductivity in this system should disappear. However, since 
we are interested in the small carrier concentrations, 
it is assumed that this relation does not take place.

As it will be shown below such a dependence of $r_{0}$
leads to a qualitatively different behavior of the system with
respect to the case with $r_{0}(n_{f})=const$. In particular,
there exists a minimal value for the coupling constant when
the two-particle bound states exist at low carrier densities
in the $s$-wave pairing channel. The existence of such 
a threshold value of the coupling constant is typical
for the $d$-wave pairing case (see, for example \cite{Duncan,Kagan}).
Another interesting property is: for any coupling constant 
in both channels there exists a corresponding carrier density value, 
below which the system is in the superconducting state.

\section{THE MODEL AND THE MAIN EQUATIONS}

The Hamiltonian of the system which describes the non-retarded fermion 
interaction can be written in a standard form
\begin{eqnarray} 
H=-\sum_{\sigma =\uparrow ,\downarrow}\int dx\psi_{\sigma}^{\dagger}(x)(\frac{\nabla^2}{2m}+\mu)\psi_{\sigma} (x)+\nonumber \\  
\int\int dx_1dx_2{\tilde V}(x_1,x_2)\psi_{\uparrow}^{\dagger}(x_1)\psi_{\downarrow}^{\dagger}(x_2)\psi_{\downarrow} (x_2)\psi_{\uparrow} (x_1),
\label{Hamilton} 
\end{eqnarray} 
where $m$ is the effective fermion mass, and $\mu$ is
the chemical potential; fermi-operators $\psi_{\sigma}(x)$ depend
on the space-time coordinate $x=({\bf x},t)$. 
The instantaneous interaction potential is chosen in the next form
$$
{\tilde V}(x_{1},x_{2})=\delta (t_{1}-t_{2})V(r),
$$
with
\begin{equation}
V(r)=V_{rep}\theta (r_1 -r)-V_{attr}\theta (r-r_1)\theta (r_0 -r),
\label{magnoninteraction}
\end{equation} 
which corresponds to potential used in \cite{Duncan}.
Here $r=|{\bf x}_1-{\bf x}_2|$ is an inter-particle distance.
Positive parameters $V_{rep}$ and $V_{attr}$ 
correspond to particle repulsion
at $r<r_1$ and particle attraction at $r_1 < r < r_0$.
The charge carrier density dependence of the correlation radius $r_0$ 
is given by (\ref{R0})
with the parameter $a =\sqrt{2/\pi}a_0$, $a_0$ is the square lattice constant.
This relation can be easily estimated from the equality
$(\pi /2) r_{0}^{2}N_{f}=a_{0}^{2}N_{cell}$, 
where on the left side the volume of the 2D system is expressed as a
volume (circle of the radius $\sim r_{0}$) occupied by one particle,
multiplied by the full number of particles $N_{f}$, 
$N^{cell}$ is an elementary cell number in the system.
The free fermion bandwidth $W$ is connected with $a_{0}$ as
$W=\pi^{2}/(ma_{0}^{2})$.
It should be noted, that the relation (\ref{R0}) at $a=\sqrt{2/\pi}a_{0}$ 
is in a good agreement with the experimental data for
$La_{2-x}Sr_{x}CuO_{4}$ \cite{Thurston}, where the plane
magnetically ordered lattice parameters are equal 
to $5.354$\AA \ \  and $5.401$\AA , 
 \ \ and the corresponding parameter $a$  is $\simeq 3.8$\AA.

In order to study the superconducting properties of the model
in the channels with different pair angular momentum $l$,
it is convenient, similarly to \cite{Duncan}, to approximate the Fourier 
transform of (\ref{magnoninteraction}) by a separable potential:
\begin{equation}
V_{{\bf k}_1{\bf k}_2}^{l}=-\lambda_{l}w_{l}({\bf k}_1)w_{l}({\bf k}_2),
\label{potentialk}
\end{equation}
where $\lambda_l$ is an effective coupling constant, and functions
\begin{equation}
w_{l}({\bf k})=h_{l}(k)cos(l\varphi_{\bf k})
\label{w}
\end{equation}
with coefficients
\begin{equation}
h_{l}(k)=\frac{(k/k_1)^l}{(1+k/k_0)^{l+1/2}},
\label{h}
\end{equation}
$k$ is the momentum modulus $k=|{\bf k}|$, and $\varphi_{\bf k}$ is its 
angle in polar coordinates
${\bf k}=(k cos(\varphi_{\bf k}), k sin(\varphi_{\bf k}))$.
Parameters $k_0$ and $k_1$ put the momentum range in the proper
region. They are
connected with the potential (\ref{magnoninteraction}) parameters
as $k_0\sim 1/r_0$ and $k_1\sim 1/r_1$. Below we put 
$k_0=1/r_0$ and $k_1 = 1/r_1$. Obviously, the expression
(\ref{potentialk}) is not the exact Fourier transform of
(\ref{magnoninteraction}), but it sets the interaction in right
momentum range and has the correct asymptotic behavior at small and large 
momenta: $V_{{\bf k}_1{\bf k}_2}^{l}\sim k_1^lk_2^l$ and 
$V_{{\bf k}_1{\bf k}_2}^{l}\sim 1/\sqrt{k_1k_2}$, correspondingly.
Since we are mainly interested in the low carrier density region,
where the crossover can take place, the correct behavior of the
interaction potential at small momenta is the most important.
In the case of low carrier concentrations the small momenta give
the main contribution in the integral of the gap equation 
(see Eq. (\ref{gapequation}) below).
We shall study $s$- and $d$-wave channels with $l=0$ and $2$ separately, so
we assume that the parameters $\lambda_l$ for both channels
are independent. It will be assumed also that the pairing takes place
at zero pair momentum  at $T=0$.

The minimization of the ground state energy with respect 
to the superconducting order parameter 
$\Delta_{l} (x_1-x_2)=V(x_1-x_2)<\psi_{\uparrow}(x_1)\psi_{\downarrow}(x_2)>$ 
and the chemical potential
gives in the case of the approximation (\ref{potentialk}) the standard 
pair of equation for the crossover problem
(see, for example \cite{Duncan}):
\begin{equation}
\Delta_{l} ({\bf k}) =-\lambda_{l}\int\frac{d {\bf p}}{(2\pi)^{2}}
\frac{\Delta_{l} ({\bf p})}{2\sqrt{\varepsilon^{2}({\bf p})+\Delta_{l}({\bf p})^{2}}}
V_{{\bf p},{\bf k}}^{l}, 
\label{gapequation}
\end{equation}
\begin{equation}
\int\frac{d {\bf p}}{(2\pi)^{2}}
(1-\frac{\varepsilon ({\bf p})}{\sqrt{\varepsilon^{2}({\bf p})+\Delta_{l}({\bf p})^{2}}}) 
=\frac{n_{f}}{a_{0}^{2}},
\label{numberequation}
\end{equation}
where $\varepsilon({\bf p})={\bf p}^{2}/2m-\mu$ 
is a free particle dispersion law.
The form of the equation (\ref{gapequation}) allows to search the
solution for the superconducting order parameter in the next form
\begin{equation}
\Delta_{l} ({\bf k}) =\Delta_{0l} w_{l}({\bf k}),
\label{gap}
\end{equation}
where $\Delta_{0l}$ does not depend on momentum ${\bf k}$
and $w_{l}({\bf k})$ is defined by (\ref{w}) and (\ref{h}).
In this case the solution of the system 
(\ref{gapequation}),(\ref{numberequation})
gives the dependence 

\begin{figure}
\begin{center}
\epsfig{file=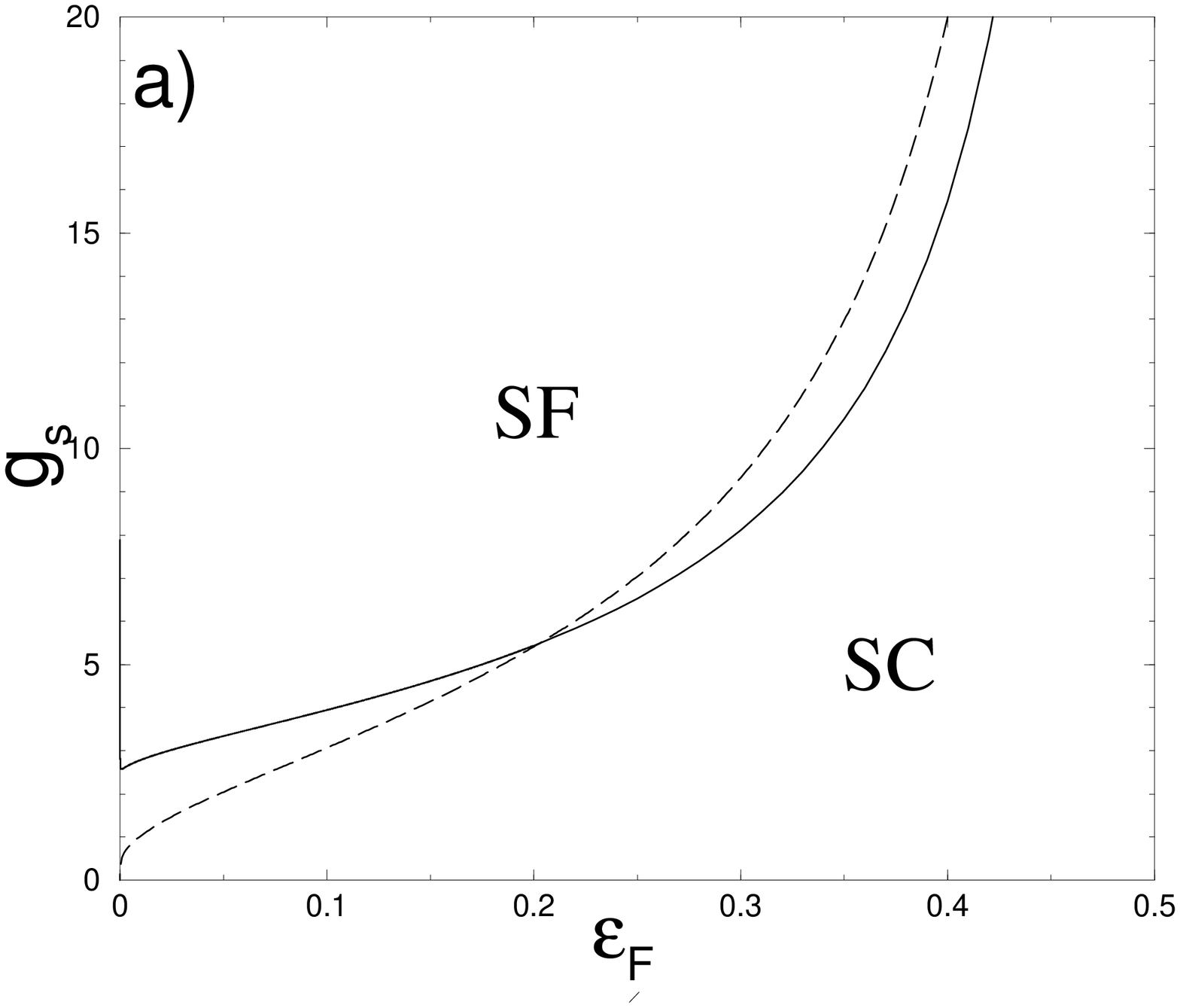,height=3.0in,width=3.0in,angle=270}
\epsfig{file=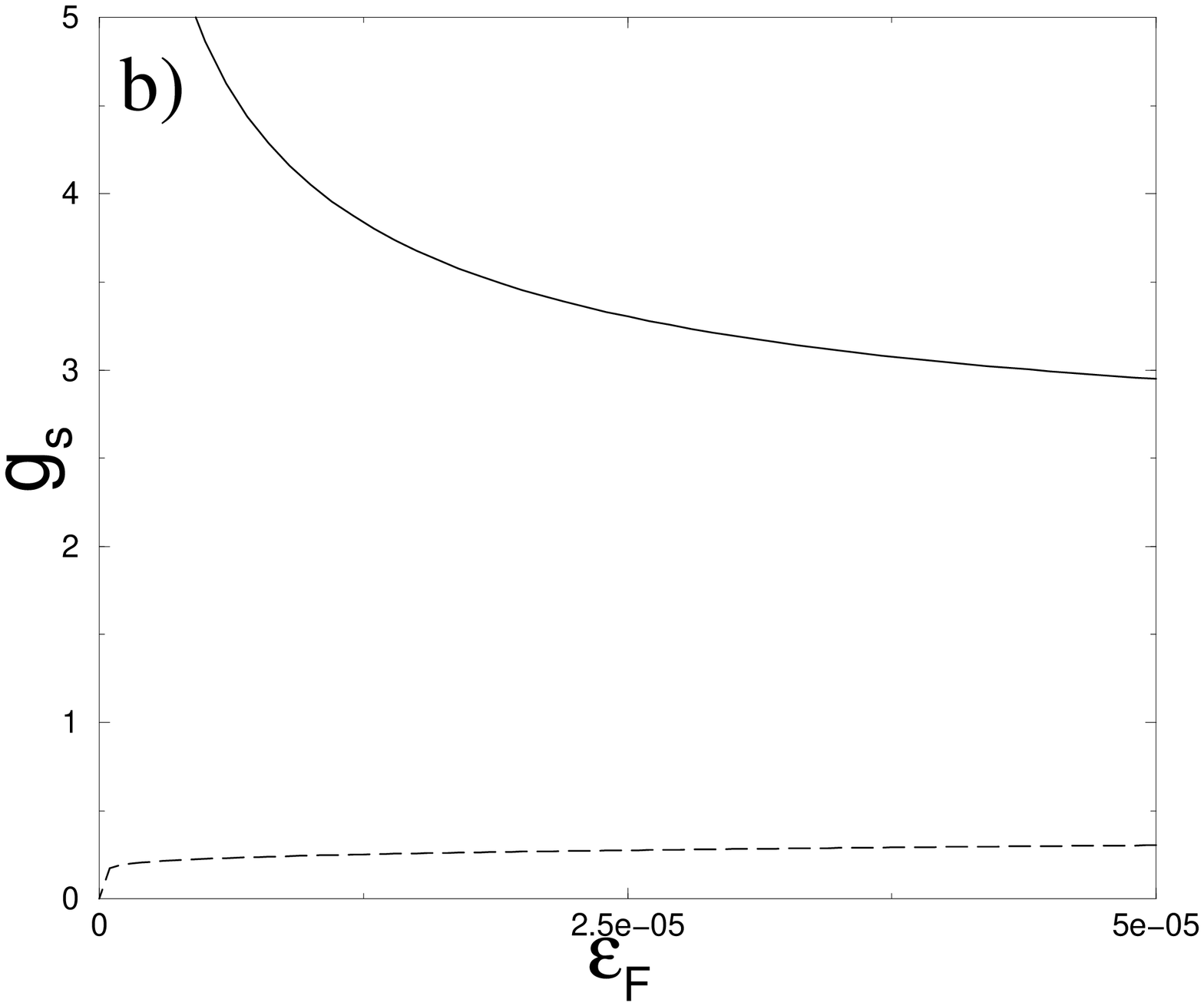,height=3.0in,width=3.0in,angle=270}
\end{center}
\caption{
a) Crossover line $g_{s}-\varepsilon_F$ 
is presented for the $s$-pairing channel
(solid line). The dotted line represents the corresponding curve
for the case with $r_{0}(n_{f})=const$ at $r_{0}=a_{0}$.
Here and below all parameters are expressed in units of the bandwidth $W$.
b) The same as a) at very low charge carrier densities.}
\label{foobar:fig1}
\end{figure}

of the gap parameter $\Delta_{0l}$
($\Delta_{0s}$ for $l=0$ and $\Delta_{0d}$ for $l=2$) 
and the corresponding chemical potential $\mu$ 
on the particle density $n_{f}$ and the coupling $\lambda_{l}$.

\section{CROSSOVER FROM SUPERFLUIDITY TO SUPERCONDUCTIVITY}

\subsection{$s$-wave pairing channel}

The solution of the system (\ref{gapequation}),(\ref{numberequation})
at $\mu =0$ gives the crossover line $\lambda_s (n_f)$, which
separates the parameters regions, where 
the local ($\mu <0$) and Cooper pairing ($\mu >0$) take place.
Note, that in the  $s$-wave pairing case the Coulomb repulsion 
parameter $k_1$ does not enter in the equations.
The numerical solution for the crossover line for the $s$-wave 
pairing case is presented in Fig.1, where
we have put Fermi energy $\varepsilon_F$ instead of $n_f$, since in the 2D case
they are connected by a linear relation $\varepsilon_F =\pi n_f/(m a_0^2)$,
and the dimensionless coupling constant $g_{s}=m\lambda_{s}/(4\pi)$
is introduced. The solution for the $r_{0}(n_{f})=const$ case is
also presented.
As it follows from the numerical calculations,
there exists a minimal value
of coupling $g_s=\simeq 2.574$, which corresponds to carrier density
$\varepsilon_F\simeq 6.8\times 10^{-4}W$ on the crossover line, necessary
to generate the crossover from superfluidity to superconductivity
with doping, otherwise the Cooper pairing regime takes place at any 
charge carrier density. In other words, there is
a minimal value of coupling constant which leads to the 
two-particle bound states in the $s$-wave pairing channel at small 
$\varepsilon_{F}$.
It is important to note, that there is no such a minimal coupling 
in the $s$-wave pairing channel when the correlation length 
$r_{0}(n_{f})=const$ \cite{Duncan}. Moreover, the crossover with charge
carrier density increasing in the
$s$-wave channel takes place for any known 
doping independent interaction potential \cite{PhysRep}.
 
Another interesting property, which follows from the Fig.1b), is the
``inverse'' crossover from superfluidity to superconductivity 
with charge carrier density decreasing at small values of 
$\varepsilon_F$ ($\varepsilon_F < 6.8\times 10^{-4} W$).
This is a consequence of the competition between
two opposite processes which occur with $\epsilon_{F}$ decreasing.
The lowering of $\varepsilon_F$ 
tends the system to become a superfluid, but at the same time it leads
to growing of $r_{0}$ and 

\begin{figure}
\begin{center}
\epsfig{file=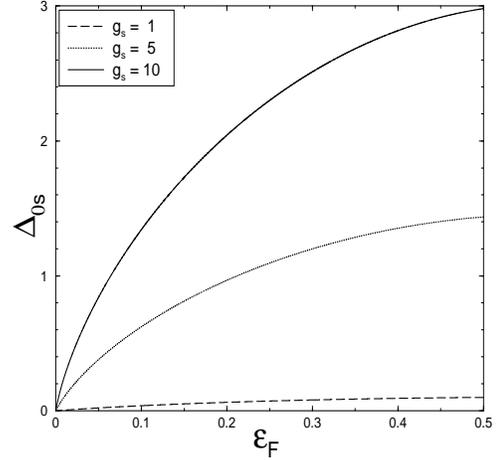,height=2.5in,width=2.5in,angle=270}
\end{center}
\caption{The dependence of $\Delta_{0s}$ on $\varepsilon_{F}$ at different
coupling parameters $g_{s}$ is presented for the $s$-wave case.}
\label{foobar:fig2}
\end{figure}

\begin{figure}
\begin{center}
\epsfig{file=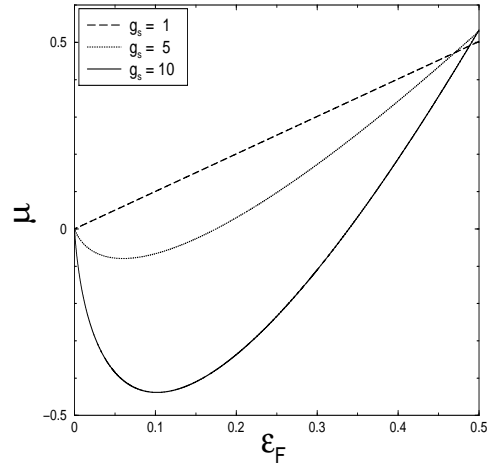,height=2.5in,width=2.5in,angle=270}
\end{center}
\caption{The dependence of $\mu$ on $\varepsilon_{F}$ 
is presented for the $s$-wave pairing channel at different values of $g_{s}$.}
\label{foobar:fig3}
\end{figure}

\begin{figure}
\begin{center}
\epsfig{file=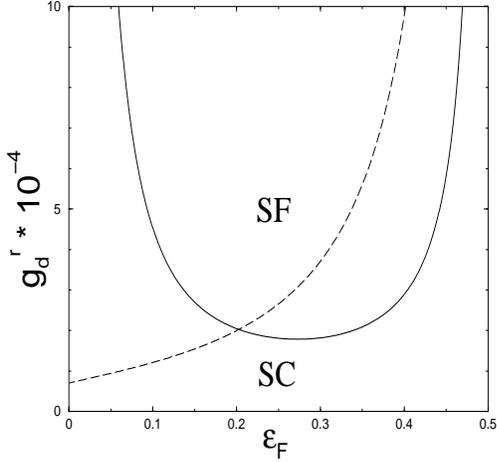,height=2.5in,width=2.5in,angle=270}
\end{center}
\caption{The crossover line $g_d^{r}(\varepsilon_{F})$ 
is presented for the $d$-wave case (solid line). The dotted line is the
crossover curve for the case $r_{0}(n_{f})=const$ at $r_{0}=a_{0}$.}
\label{foobar:fig4}
\end{figure}

makes the pair size larger, i.e. the pairs become bounded weaker.
On the other hand,
the carrier localization on dopants at very small carrier densities
also makes a possible density region of superfluidity more narrow
\cite{Pogorelov}.
However, the region of extremely small particle densities
$\varepsilon_F\rightarrow 0$ is not very interesting from the point
of view of the connection with HTSC, since in this region the relation 
$r_{0}\sim 1/\sqrt{n_{f}}$
does not hold, and the model is not correct. 
As it will be shown in the next Subsection, in the $d$-wave case
this crossover from from superfluidity to superconductivity with
doping decreasing takes place at rather large Fermi energy values.

The doping dependencies of the gap $\Delta_{0s}$ and the chemical potential
$\mu$ at different values of the coupling constant $g_{s}$
are presented in Figs. 2, and 3. 
The gap is increasing with the doping almost linearly,
except the region of extremely low concentrations. It is interesting 
to note, that the chemical potential in the superfluidity region ($\mu <0$) 
has a minimum at a finite value of $\varepsilon_{F}$
and it is equal to zero at $\varepsilon_{F}=0$, since the effective
coupling constant at low doping 
\begin{equation}
g_{l}\sim k_{0} k_{F}^{2l}g_{l}\sim\varepsilon_{F}^{l+1/2} 
\label{renormg}
\end{equation}
is zero at $\varepsilon_{F}=0$.
In other words, the local pairs are the most strongly coupled at some finite
carrier density value. This is also a consequence of the competition
between the Fermi surface
formation and correlation radius decreasing with charge carrier density growth.
This situation is qualitatively different from the $r_{0}(n_{f})=const$
case, where $k_{0}$ does not depend on particle density and renormalized
coupling constant is $\varepsilon_{F}$-independent
in the $s$-channel. In this case
$\mu =E_{b}/2<0$ at $\varepsilon_{F}=0$,
where $E_{b}<0$ is a two-particle bound state energy.  

\subsection{$d$-wave pairing channel}

In this case the Coulomb repulsion parameter $k_{1}$ is present
in the equations (\ref{gapequation}) and (\ref{numberequation}).
However, the presence of this parameter leads just to renormalization of
the dimensionless coupling constant $g_d =m\lambda_d/(4\pi)$ 
and the energy
gap parameter: $g_d^{r}=g_d/(\varepsilon_{1}/W)^{2}$, 
$\Delta_{0d}^{r}=\Delta_{0d}/\varepsilon_{1}$, where 
$\varepsilon_{1}=k_{1}^{2}/(2m)$ is characteristic energy of Coulomb
repulsion. We shall consider the case, when the Coulomb repulsion
is much smaller than $W$, i.e. 
this is the case of a large free-fermion 
bandwidth. Therefore, the coupling constant
is assumed to be large: $g_{d}^{r}\sim 10^{4}-10^{5}$ (see Figs. 4-6). 
The crossover line $g_{d}^{r}(\epsilon_F)$ 
for the $d$-wave pairing case is presented in Fig.4.
Qualitatively, the behavior of the system with doping and coupling changing
in the $d$-wave pairing channel is similar to the $s$-wave case.
The important difference is that the low carrier density superconducting
state exists at rather high values of $\varepsilon_{F}$.
Also in this case there is a minimal value of the coupling
constant for two-particle bound states $g_{d}^{r}\simeq 1.7856\times
10^{4}$ at $\varepsilon_{F}\simeq 0.2731 W$.
It should be noted that the existence of the large threshold value for
the coupling constant in both channels can be a possible answer
on the question why the crossover has been not observed in cuprates.

The charge carrier density dependence of $\Delta_{0d}^{r}$ and $\mu$ 
at different coupling parameters $g_{d}^{r}$ are presented
in Figs.5 and 6. The superconducting gap $\Delta_{0d}$ at low charge
carrier densities
is much smaller than in the $s$-case, due to stronger 
$\varepsilon_{F}$-dependence of effective coupling 
constant at low carrier densities (\ref{renormg}).
The magnitude of the order parameter $\Delta_{0d}^{r}$ 
starts to grow almost linearly with $\varepsilon_{F}$
increasing when the Fermi energy is larger of some minimal value (Fig.5). 
This behavior is qualitatively similar to the doping dependence
of the gap of cuprates in the underdoped regime. However, because of
its simplicity, the model
can't describe the decreasing of the gap with charge carrier
density increasing at large values of $n_{f}$ HTSCs. For this
other properties of the charge carrier interaction 
in cuprates have to be taken into account (see the last Section). 

The small-$\varepsilon_{F}$ superconducting region with
$\mu > 0$ as well as region of superfluidity are
rather large in the $d$-wave pairing channel. 
Also in this case the chemical potential
is equal to zero at $\varepsilon_{F}=0$, i.e. there are no two-particle
bound states at very low charge carrier densities. 
It should be noted that the decreasing of the chemical potential
with increasing charge carrier density at small $\varepsilon_{F}$ 
both in $s$- and $d$-wave pairing
cases indicates a negative electronic
compressibility. This can be related 
to increasing of antiferromagnetic 
correlations at low carrier densities.

\section{CONCLUSION}

Theoretical description of the behavior of superconductor 
with carrier density changing is an interesting and important problem, 
in particular, because of its possible association with HTSCs. 
The microscopic mechanism of the superconductivity in cuprates is not 
known so far, and the solution of phenomenological models, which take

\begin{figure}
\begin{center}
\epsfig{file=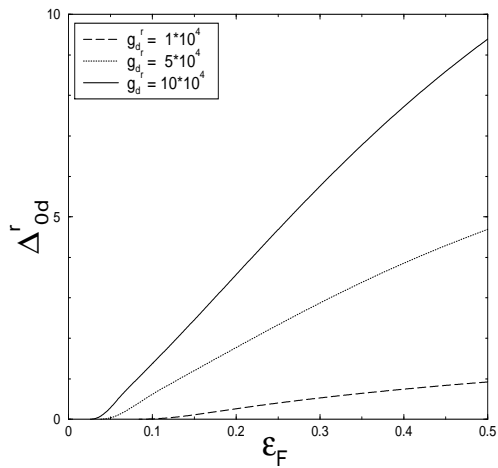,height=2.5in,width=2.5in,angle=270}
\end{center}
\caption{The dependence of $\Delta_{0d}^{r}$ on $\varepsilon_{F}$ is presented for the 
$d$-wave pairing channel at different values of coupling constant $g_{d}^{r}$.}
\label{foobar:fig5}
\end{figure}

\begin{figure}
\begin{center}
\epsfig{file=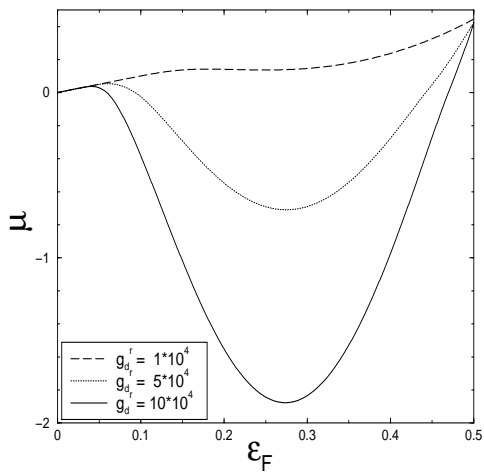,height=2.5in,width=2.5in,angle=270}
\end{center}
\caption{Chemical potential $\mu$ as a function of $\varepsilon_{F}$ is presented 
for the $d$-wave case at different values of coupling constant $g_{d}^{r}$.}
\label{foobar:fig6}
\end{figure}
 
into account some of the properties of HTSCs, can help to clarify 
the nature of their unusual behavior, and maybe even help to understand 
the microscopic mechanism of the HTSC phenomenon.

In this paper the possibility of the crossover
from superfluidity to superconductivity with charge carrier density
and coupling constant changing in different pairing channels at $T=0$
was studied for a model, 
which qualitatively takes into account one of the properties of HTSCs, namely,
the doping dependence of correlation length $r_0$ at low carrier densities.
It has been shown, that even this simple model results in interesting 
and unusual properties, which are rather different from a more standard 
case with $r_0(n_{f})=const$. In particular, the two-particle bound states 
in the $s$-channel exist only if the coupling constant is larger of the
threshold value, similar to the $d$-wave pairing case. 
At any value of coupling constant larger of the threshold one,
the ``inverse'' crossover from superfluidity to superconductivity takes place
with doping decreasing in both $s$- and $d$-wave channels.
Of course, such a simple model can not pretend to describe doping dependence
of the gap and chemical potential of HTSCs. 
The momentum dependence of the interaction potential
has to be taken into account more carefully, especially in the overdoped
regime, where $\varepsilon_F$ is rather large and the separable potential 
may be not correct. In general, also the effect of the retardation of
interaction can not be neglected. These and some other questions
are planed to be studied in a future work.

\section*{Acknowledgments}

V.T.~thanks the CFIF members, especially Prof.~P.D.~Sacramento and 
Prof.~V.R.~Vieira for kind hospitality.
V.M.L. is partly supported by SCOPES-project 7UKPJ062150.00/1 of the
Swiss National Science Fundation.

\end{document}